\documentclass[%
 reprint,
 amsmath,amssymb,
 aps,
prb,
]{revtex4-1}
\usepackage{romannum}
\usepackage[caption=false]{subfig}
\usepackage{graphicx}
\usepackage{dcolumn}
\usepackage{bm}

\begin{document}
\pagenumbering{arabic}
\setcounter{page}{1}
\preprint{APS/123-QED}

\title{Number Parity effects in the normal state of $SrTiO_3$}

\author{Xing Yang}
\affiliation{%
School of Physics and Electronics, Hunan University, Hunan, China, 410082
\\
 Physics Department, University of Notre Dame, Notre Dame, Indiana, USA, 45665
}%

\author{Quanhui Liu}
\affiliation{
 School of Physics and Electronics, Hunan University, Hunan, China, 410082
}%

\author{Boldizs\'ar Jank\'o}
\affiliation{
  Physics Department, University of Notre Dame, Notre Dame, Indiana, USA, 45665
 }%

\date{\today}

\begin{abstract}
We study the recently discovered even-odd effects in the normal state of single-electron devices manufactured at strontium titanium oxide/lanthanum aluminum oxide interfaces (STO/LAO). Within the framework of the number parity-projected formalism and a phenomenological fermion-boson model we find that, in sharp contrast to conventional superconductors, the crossover temperature $T^*$ for the onset of number parity effect is considerably larger than the superconducting transition temperature $T_c$ due to the existence of a pairing gap above $T_c$.  Furthermore, the finite lifetime of the preformed pairs reduces by several orders of magnitude the effective number of states $N_{\rm eff}$ available for the unpaired quasiparticle in the odd parity state of the Coulomb blockaded STO/LAO island. Our findings are in qualitative agreement with the experimental results reported by Levy and coworkers for STO/LAO based single electron devices.
\begin{description}
\item[PACS numbers]
73.23.Hk, 73.40.-c, 73.63.-b, 74.25.-q, 74.78.-w

\end{description}
\end{abstract}

\pacs{73.23.Hk, 73.40.-c, 73.63.-b, 74.25.-q, 74.78.-w}
\maketitle
\makeatletter
\newcommand*{\rom}[1]{\expandafter\@slowromancap\romannumeral #1@}
\makeatother

\section{Introduction}

Number parity effects in superconductors were expected as soon as the Bardeen-Cooper-Schrieffer (BCS) microscopic model was developed \citep[][]{BCS}. Indeed, the BCS ground state corresponds to a coherent superposition of {\em pair} states in which the number of particles has even parity and the total number $N$ is not fixed. Under these circumstances, the charge displacement operator $exp( i \phi)$ (canonically conjugate with the number operator $\hat N$) has a fixed expectation value, which leads to the common notion that a macroscopic BCS superconductor has a complex order parameter $\Delta$ with a rigid phase $\phi$. As soon as the BCS state is projected \cite{Schrieffer} onto {\em fixed N}, it becomes clear that one has to differentiate between two cases: (a) if the total number $N= 2n$ is even, all particles can participate in pair states and the ground state resembles the usual grand canonical BCS ground state; (b) if $N=2n+1$ is odd, the ground state will inevitably contain not only pairs, but also an unpaired electron (more precisely: the ground state will contain a Bogoliubov quasiparticle).

Intuitively, one would expect that the $N$ vs $N+1$ (even/odd) difference in a superconductor or any kind of paired fermionic state must be experimentally observable only if N is relatively small. Indeed, inspired by the success of the BCS theory, Bohr, Mottelson and Pines \cite{Bohr:1958aa} were the first of many who studied pairing and even-odd effects in nuclear matter, with particle numbers around $ N \sim 10^2$. It was therefore, even more surprising, when Mooij et al. \cite{Geerligs:1990aa}, Tinkham and coworkers \cite{Tuominen:1992aa}, as well as  Devoret and his colleagues \cite{Lafarge:1993aa,Lafarge:1993ab} showed experimentally measurable difference between Coulomb blockaded mesoscopic superconducting islands that contain a billion, and a billion plus one electrons. As it turns out, the magnitude of $N$ was less important. Instead, the quality of the Coulomb blockade turned out to be crucial: the superconducting islands had to be isolated from their environment with ultrasmall tunnel junctions and highly resistive electromagnetic environment, in order to ensure that $N$ is a fixed, good quantum number.

The pioneering experiments on number parity effects in conventional superconductors were performed on single-electron (SET) devices consisting of lithographically patterned aluminum islands \cite{Grabert1992}. Even-odd effects emerged below a crossover temperature $T^*$ that was always much lower than the superconducting transition temperature: $T^* \ll T_c$. Rather than being directly correlated with $T_c$, $T^*$ is set by the experimentally measurable even-odd free energy difference $\delta {\cal F}_{\rm e/ o} \sim \Delta_0 - k_B T \log N_{\rm eff}$. Here $\Delta_0$ is the low temperature energy gap, and $N_{\rm eff}$ is the effective number of states \cite{Tuominen:1992aa,Lafarge:1993aa,Lafarge:1993ab,Janko:1994aa} available for the unpaired electron to explore in the odd number parity state of the superconducting island. Within this parity projected framework \cite{Tuominen:1992aa,Janko:1994aa} $T^*$ corresponds to the temperature at which $\delta {\cal F}_{\rm e/ o}$ becomes negligibly small: $ T^* \sim \Delta_0/(k_B \log N_{\rm eff})$. For typical device parameters in these early experiments, the crossover temperature was measured to be around $T^* \sim 10^2 {\rm mK}$ for aluminum island with $T_c \sim 1 {\rm K}$. Consequently, the effective number of states was typically around $N_{\rm eff} \sim 10^4$.

The experiments by Levy and his coworkers \cite{Cheng:2015aa} on SET devices constructed on STO/LAO provided experimental evidence for a spectacular departure from the conventional number parity effects described above. Levy and his colleagues detected $T^* \sim 900 {\rm mK}$, much higher than the superconducting transition temperature $T_c \sim 300 {\rm mK}$ measured for these devices. Even-odd effects remained detectable well into the "normal" phase of the superconductor, and persisted in magnetic fields $B^* \sim  1\sim4 {\rm T}$, much higher than the upper critical field of the device. Furthermore, the extracted $N_{\rm eff}$ is also drastically different: $N_{\rm eff} \sim 2-3 $.

A possible and relatively straightforward interpretation of novel experimental developments suggest that preformed pairs \cite{GeshkenbeinSuperconductivity,Seo_2019,PhysRevB.94.155114} persist into the normal state of STO-LAO well above the superconducting transition temperature. Consequently, fundamental changes must be made to the theoretical description of the number parity effects in this novel preformed pair phase. This paper is devoted to the presentation of a phenomenological theoretical framework aimed at providing a description of number parity effects in the normal phase of STO/LAO devices. Given the fact that the details of the microscopic mechanism behind the superconducting and preformed pair state of STO/LAO are not yet established, we use a phenomenological fermion-boson model \cite{Alexandrov:1996} that allows us to describe a normal phase where both pairs and unpaired particles are present. Furthermore, the model allows pairs to decay into unpaired particles, and particles to form pairs. This theoretical picture provides in a natural way a finite pair lifetime \cite{Cheng:2015aa} in the preformed pair state. We find, after performing the number parity projection developed earlier by Ambegaokar, Smith and one of us \cite{Janko:1994aa},  that the finite pair lifetime has drastic effect on the magnitude of $N_{\rm eff}$. In fact, as we will show in detail below, the theoretical framework we develop in this paper can reproduce not only $T^* \gg T_c$, but also $N_{\rm eff} \sim \mathcal{O}(1) $.

\begin{figure}[htbp]
\subfloat[]{\includegraphics[height=1.8in]{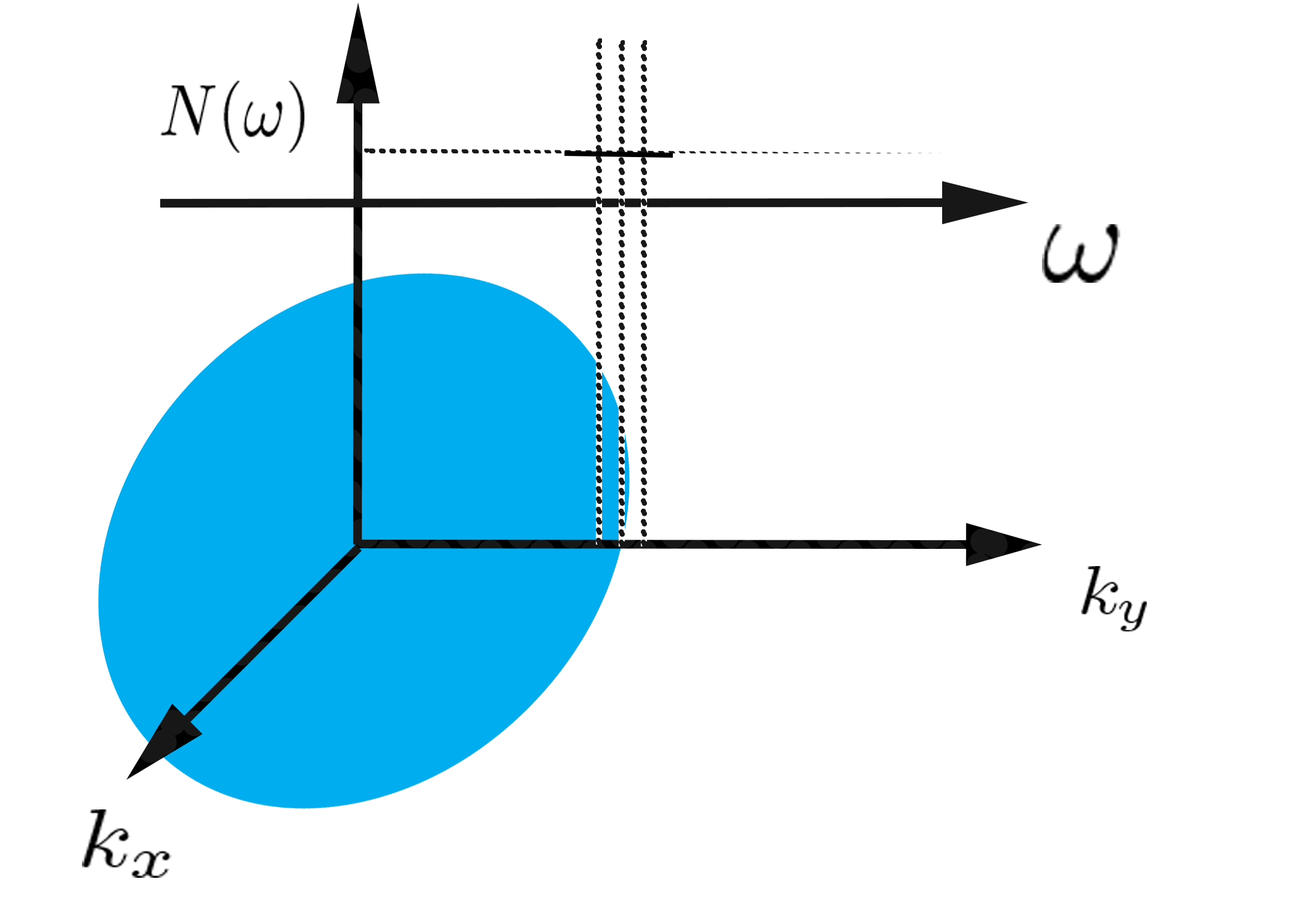}}
\\
\subfloat[]{\includegraphics[height=1.8in]{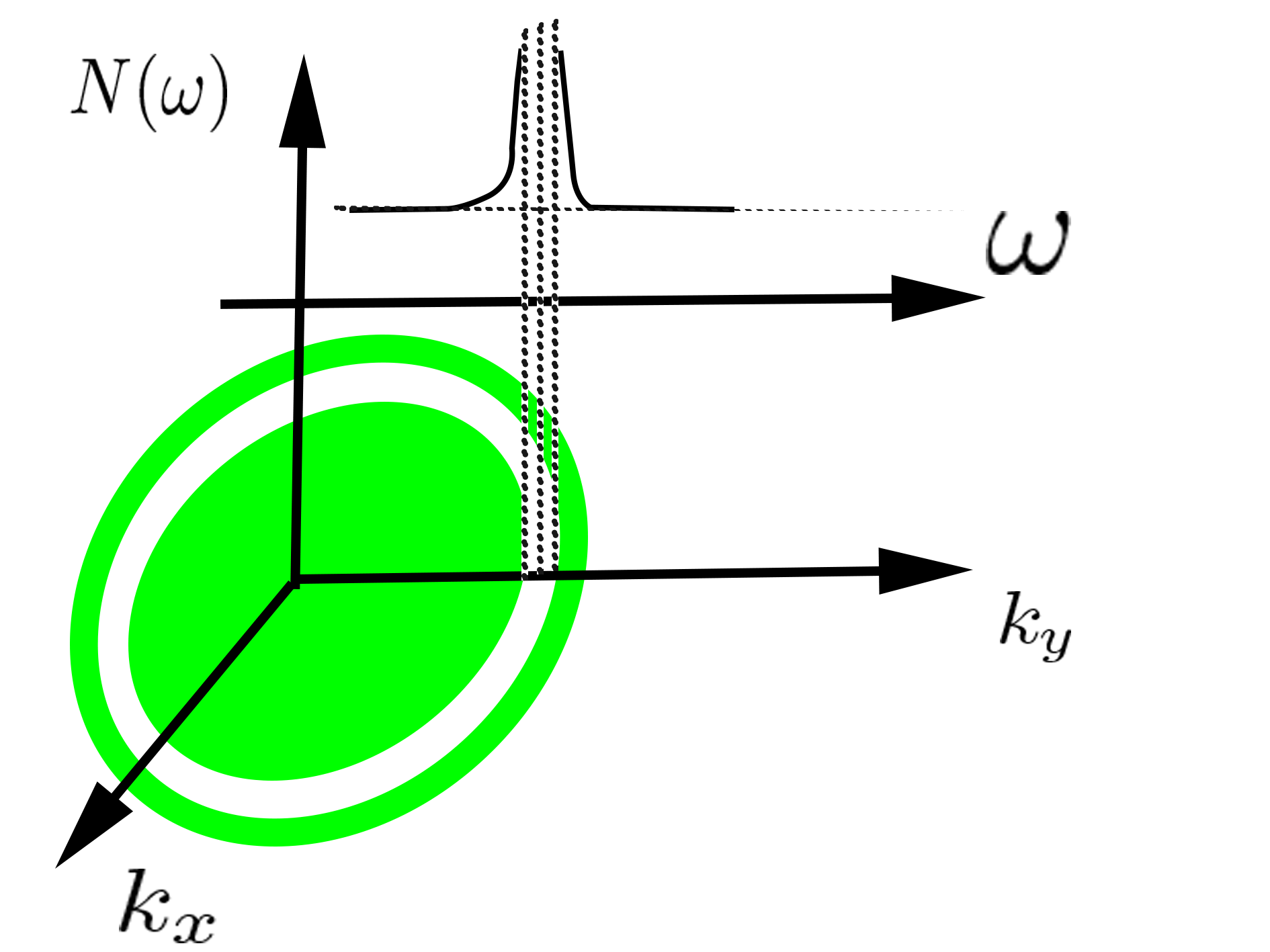}}
\\
\subfloat[]{\includegraphics[height=1.8in]{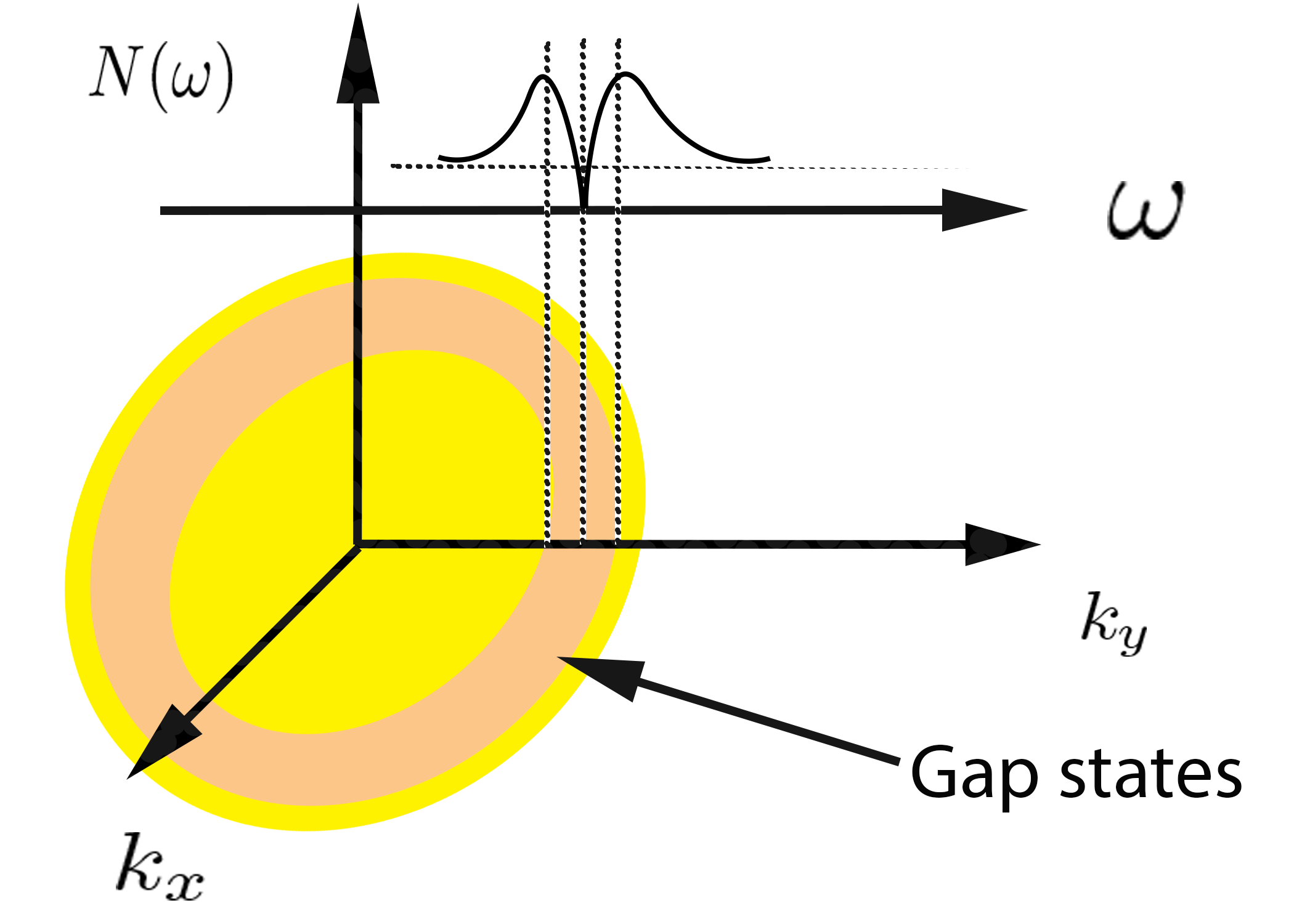}}
\caption{Qualitative sketch of the momentum space and density of states for (a) the density of states in the metallic normal state; (b) a conventional superconductor with a superconducting gap, and (c) an unconventional superconductor with a zero-width gap at Fermi level. For the cases (b) and (c), even-odd effect can be expected because of the presence of the superconducting gaps.}
\label{pic1}
\end{figure}
Generally speaking, the materials for making the single-electron devices can be separated into three categories (see Fig. \ref{pic1}): gapless materials, BCS and unconventional superconductors. Their density of states are shown, respectively, in panels (a), (b) and (c) of Fig. \ref{pic1}. While the single electron transistors with {\it ultra}small islands and discrete energy spectrum have also been investigated extensively \cite{PhysRevLett.77.4962,PhysRevLett.77.3189}, we will not discuss this regime here. According to our calculations, the density of states at the Fermi level should be vanishingly small in order to obtain a finite even/odd free energy difference. As a result, the single electron transistors made from BCS superconductors and unconventional superconductors (as shown in panels (b) and (c)) are expected to show even-odd effects, and a superconducting gap above $T_c$ is necessary to cause $T^*\gg T_c$. The effective excitation number for the unpaired electrons in the odd parity states is highly dependent on the density of states at $E=\Delta$, since the smallest excitation energy is assumed to be $\Delta$ \cite{Tuominen:1992aa, Lafarge:1993aa, Lafarge:1993ab}. In BCS superconductors (see panel (b)), the density of states at $E=\Delta$ is known to have a van Hove singularity, and this results in a large $N_{\rm eff}\sim 10^4$. In unconventional superconductors (see panel (c)), the van Hove singularity is broadened by the presence of low energy quasiparticles, which results in a small $N_{\rm eff} \sim \mathcal{O}(1)$.

Several possible microscopic superconducting mechanisms of the electron system at the STO/LAO interface have been proposed recently by different groups. Ruhman and Lee \cite{PhysRevB.94.224515} suggested on the plasmon-induced superconducting mechanism. A nonperturbative approach within the plasmon model is being developed by Edelman and Littlewood \cite{Edelman}. Kedem {\it et al.} \cite{PhysRevLett.115.247002,PhysRevB.98.220505} related the mechanism to the ferroelectric mode. Arce-Gamboa and Guzm\'an-Verri \cite{PhysRevMaterials.2.104804} discovered the influence of strain on the ferroelectric mode and obtained the dependence of superconducting transition temperature on cation doping. On the contrary, W\"olfle and Balastsky \cite{PhysRevB.98.104505} proposed the transverse optical phonons may be the glue for electron pairing. While these theories can explain the origin of superconducting gap selfconsistently, some important experimental facts in remain unexplained. First of all, the pairing gap should persist above $T_c$. This is very important for the even-odd effects above $T_c$ in the single electron transistors\cite{Richter2013}. Next, the van Hove singularity in the density of states should be broadened out. This will lead to a small $N_{\rm eff}$ consistent with the experiments of Levy {\it et al} \cite{Cheng:2015aa}. The detailed discussion of these results will be presented in the sections below.

This paper is organized as follows. In Section \ref{sec:phq}, two important physical quantities, the even/odd free energy difference and the effective excitation number for the unpaired electron in the odd parity state, are related to the density of states within the phenomenological Dynes formula. In Section \ref{sec: bf}, the boson-fermion model is introduced, and the analytic form of its electron Green's function is provided. The density of states and the physical quantities of the even-odd effect predicted by the model are calculated in Section \ref{sec:dos}.  Finally, we present our conclusions in Section \ref{sec:con}.

\section{even/odd free energy difference and effective excitation number for the unpaired electron}
\label{sec:phq}
\subsection{Even/odd free energy difference}
The electron system of the quantum dot at the STO/LAO interface is described by a general Hamiltonian $\hat{H}$. Within the number parity projection formalism \cite{Janko:1994aa}, the canonical partition function with even/odd number parity is :
\begin{equation}
\label{eq1}
Z_{e/o}={\rm Tr} \bigg \{ {\frac{1\pm (-1)^{\hat{N}}}{2}}e^{-\beta (\hat{H}-\mu \hat{N})} \bigg \}
\end{equation}
\noindent where the symbol $e/o$ corresponds to the even/odd parity, $\hat{N}$ is the electron number operator, while $\beta = \frac{1}{k_B T}$ where $k_B$ is Boltzmann constant and $T$ is the temperature. 

From Eq. \ref{eq1}, the difference between the free energy of a system with an odd and even number of particles is 
\begin{equation}\label{eq2}
F_o -F_e=\frac{1}{\beta}\ln\bigg[\frac{1+\langle(-1)^{\hat{N}}\rangle}{1-\langle(-1)^{\hat{N}}\rangle}\bigg],
\end{equation}
\noindent where $\langle...\rangle \equiv {\rm Tr}\{e^{\beta (\hat{H}-\mu \hat{N})}...\}/Z$ and $Z={\rm Tr}  \{ e^{-\beta (\hat{H}-\mu \hat{N})}  \}$. The expectation value $\langle(-1)^{\hat{N}}\rangle$ is the parameter that signals the presence or absence of even-odd effects. When $\langle(-1)^{\hat{N}}\rangle=0$, even-odd effects will not be observable. Let us assume that the Hamiltonian can be expressed in a more compact form $\hat{H}=\sum_{\mathbf{k},\sigma}{e_\mathbf{k} \hat{c}^\dagger_{\mathbf{k},\sigma} \hat{c}_{\mathbf{k},\sigma}}$. From the above, $\langle(-1)^{\hat{N}}\rangle$ is
\begin{equation}
\label{eq3}
 \langle (-1)^{\hat{N}}\rangle=\prod_{\mathbf{k}} \frac{(1-e^{\beta (e_\mathbf{k}-\mu)})^2}{(1+e^{\beta (e_\mathbf{k}-\mu)})^2}=\prod_{\mathbf{k}} \tanh^2(\frac{\beta (e_\mathbf{k}-\mu)}{2}).
\end{equation}
If $A$ is defined as $e^A \equiv \langle (-1)^{\hat{N}}\rangle$, then
\begin{equation}
\label{eq4}
A=2 \sum_\mathbf{k} \ln \bigg | \tanh \frac{\beta e_\mathbf{k}}{2}\bigg|=2 \int^{+ \infty}_{- \infty} D(E) \ln \bigg |\tanh \frac{\beta E}{2}\bigg |  d E \\
\end{equation}
where $D(E)$ is the density of states. Notice that the factor $ \ln |\tanh \frac{\beta E}{2}|$ in the integrand is divergent when $E=0$. This suggests that an energy gap is necessary for a system to show even-odd effects. In the absence of a pairing gap $A$ is a large negative number and $e^A \approx 0$. At the interface of STO/LAO, the even-odd effects appear above the superconducting transition temperature $T_c$. This implies the existence of a pairing gap above $T_c$. Scanning tunneling spectroscopy experiments also show that an energy gap persists above $T_c$\cite{Richter2013}. It is one of the requirements for a system showing number-parity effects. Moreover, the factor $ \ln |\tanh \frac{\beta E}{2}|$ turns to be zero when $E \gg E_F$, this suggests the part with the high energy does not contribute to the integral $A$. On the other hand, the density of states near Fermi level (gap states) can increase the value of $|A|$ greatly, and the even-odd effect parameter $\langle (-1)^{\hat{N}}\rangle=e^A=e^{-|A|}\ll 1$ is reduced accordingly. In this sense, the emergence of the gap states can weaken even-odd effects.

\subsection{Effective excitation number for the unpaired electron}

With the assumption that the smallest excitation energy for electrons is $\Delta$, we can calculate the effective excitation number for the unpaired electron with the formula \cite{Tuominen:1992aa, Lafarge:1993aa, Lafarge:1993ab}:
\begin{equation}
\label{eq6}
N_{\rm eff}=\int^{\infty}_{\Delta_0}D(E)\exp(-\beta(E-\Delta)) d E.
\end{equation}

In Eq. \ref{eq6}, the density of states at $E=\Delta$ contributes most to the effective excitation number for the unpaired electron in the odd parity states. If it is assumed that the van Hove singularity in $D(E)$ exists, it can easily produce a large $N_{\rm eff} \sim 10^4$ or more in BCS superconductors. However, the experiments \cite{Cheng:2015aa} at the interface of STO/LAO discover a very small $N_{\rm eff} \sim 1$. This indicates that the van Hove singularity was broadened out in the density of states.

\subsection{Even-odd effect phenomenology with the Dynes formula}

As we can see in the above discussion, the even-odd effects are related to the density of states of the small island in the single electron transistors. In order to reproduce the experimental results, the van Hove singularity should, at least, be broadened. This can be provided by the lifetime effects of electron pairs. Here we adopt the phenomenological Dynes formula and calculate the even/odd free energy difference $\delta F_{e/o}$ and the effective excitation number for the unpaired electron $N_{\rm eff}$

\begin{figure}
  \includegraphics[width=\linewidth]{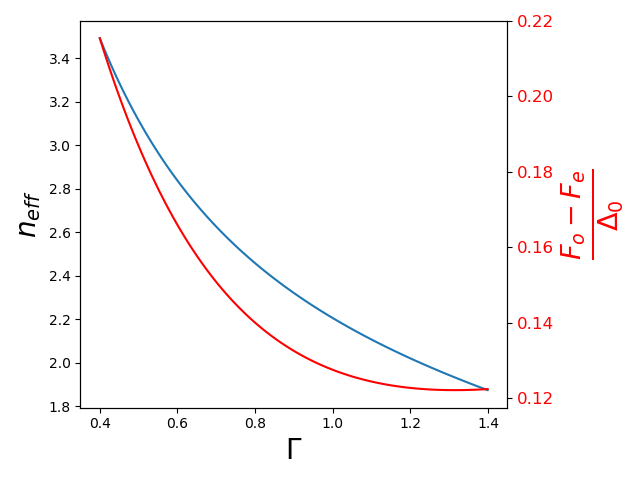}
  \caption{$\Gamma$ versus the even/odd free energy difference $\delta F_{e/o}$ and the number of the effective excitation states for the unpaired quasiparticles, $N_{\rm eff}$.}
  \label{neff}
\end{figure}

To be explicit, we will use the following form for the density of states\cite{Dynes:1978aa}

\begin{equation}\label{eq5}
D_d(E)=D_n(0) \mathrm{Re} \frac{|E- i \Gamma|}{\sqrt{(E-i \Gamma)^2 - \Delta^2}}
\end{equation}

\noindent where $D_n(0)$ is the density of states in the normal state, $\Gamma$ is the phenomenological imaginary part of energy, and $\Delta$ is the superconducting energy gap. With Eqs. \ref{eq2}-\ref{eq6}, the even/odd free energy difference $\delta F_{e/o}$ and the effective excitation state number for the unpaired electron $N_{\rm eff}$ are calculated and plotted in Fig. \ref{neff}. As we can see, $N_{\rm eff}$ reduces to $\sim 1$, and the even/odd free energy difference is finite provided that the superconducting gap $\Delta$ does not close. This result is independent of any microscopic model. In order for the pairing-induced even-odd effect to be visible, the density of states at the Fermi level must vanish, and broadening $\Gamma$ has to be small compared to the gap $\Delta_0$

\section{ The boson-fermion Model}
\label{sec: bf}
As mentioned in the introduction, there is no consensus yet on the microscopic theory of superconductivity in STO. In order to reproduce a Dynes-like density of states, we turn to the phenomenological boson-fermion model. For a single band model, electrons are assumed to have a Bogoliubov quasiparticle dispersion $E_\mathbf{k}=\sqrt{\epsilon^2_\mathbf{k}+\Delta^2}$, where $\epsilon_\mathbf{k}=\frac{\hbar^2 k^2}{2 m^*}-\mu_F$. The Hamiltonian of the electrons can be written as:
\begin{equation}
\hat{H}_{0e}=\sum_{\mathbf{k},\sigma}{E_\mathbf{k} \hat{c}^\dagger_{\mathbf{k},\sigma} \hat{c}_{\mathbf{k},\sigma}}.
\end{equation}

The superconducting gap of the 2D electron system at the interface of STO/LAO, $\Delta$ vanishes at  $T_s \sim 300 {\rm mK}$, and it turns into superconducting state at $T_c \sim 190 {\rm mK}$ \cite{Richter2013}. This suggests between $T_c$ and $T_s$, the superconducting phase is destroyed, but the superconducting gap is preserved. This regime corresponds to the preformed pair state. The present model is devoted to studying the preformed pair state and superconducting state. Notice that the coherence length of pairs is $\sim {\rm 70- 100 nm}$ in (001)-STO/LAO and ${\rm 40-75 nm}$ for (011)-STO/LAO \cite{Herranz:2015aa}, which is very small compared to the coherence length in conventional superconductors. In order to give an approximate description of preformed pairs, we introduce a bosonic field $\hat{b}_\mathbf{q}$ with elementary charge unit $-2e$. For a small momentum $\mathbf{q}$, the dispersion of the pairs is approximated as $\xi_\mathbf{q}= \xi_0+\hbar v |q|-\mu_b$ \cite{Schrieffer}, and the Hamiltonian for the bare bosonic field is:

\begin{equation}
\hat{H}_{0p}=\sum_{\mathbf{q}}{\xi_\mathbf{q} \hat{b}^\dagger_{\mathbf{q}} \hat{b}_{\mathbf{q}}}
\end{equation}

\noindent where $\hat{b}^\dagger_{\mathbf{q}}$ and $\hat{b}_{\mathbf{q}}$ are defined to commute with  $\hat{c}^\dagger_{\mathbf{k},\sigma}$ and $\hat{c}_{\mathbf{k},\sigma}$. The interaction Hamiltonian between the fermions and bosons is assumed to be:

\begin{equation}
\hat{H}_1=\sum_{\mathbf{k},\mathbf{q}}{\frac{V_1(\mathbf{q})}{\sqrt{n_0}}\hat{b}^\dagger_\mathbf{q} \hat{c}_{-\mathbf{k}+\frac{\mathbf{q}}{2}\downarrow} \hat{c}_{\mathbf{k}+\frac{\mathbf{q}}{2}\uparrow}+H.c.}
\end{equation}

The total Hamiltonian is

\begin{equation}
\hat{H}=\hat{H}_0+\hat{H}_1
\end{equation}

\noindent where

\begin{equation}
\hat{H}_0=\hat{H}_{0e}+\hat{H}_{0p}.
\end{equation}

The total particle number is defined as $\hat{N}=\sum_{\mathbf{k},\sigma}{ \hat{c}^\dagger_{\mathbf{k},\sigma} \hat{c}_{\mathbf{k},\sigma}}+2 \sum_{\mathbf{q}}{ \hat{b}^\dagger_{\mathbf{q}} \hat{b}_{\mathbf{q}}}$, and it can be proven that $[\hat{N}, \hat{H}]=0$. The first order approximation of the self-energy is

\begin{equation}
\label{eq:sigma}
\begin{split}
&\Sigma(\mathbf{k}, \omega)= \frac{1}{\hbar^2}\int \frac{L |V_1(\mathbf{q})|^2 d \mathbf{q}}{2 \pi n_0} \frac{1}{\omega-\frac{(\xi_{\mathbf{q}}-E_{\mathbf{q}-\mathbf{k}})}{\hbar}+i \eta}\\
&\times \bigg (\frac{1}{e^{\beta \xi_{\mathbf{q}}}-1}+\frac{1}{e^{\beta E_{\mathbf{q}-\mathbf{k}}}+1}\bigg )\\
\end{split}
\end{equation}

\noindent where $L$ is the length of the quantum dot in the middle of the single electron transistor. Notice that the superconductivity of the STO/LAO system is considered to be one-dimensional \cite{Cheng:2015aa}, which makes our proposed theory to be one-dimensional as well. Let us now introduce a momentum dependent interaction kernel $V_1(q)$ as an example interaction that reproduces a Dynes-like density of states:

\begin{equation}
V_1=V_c\sqrt{\frac{(\xi_{\mathbf{q}}-E_{\mathbf{q}-\mathbf{k}})^2}{(\xi_{\mathbf{q}}-E_{\mathbf{q}-\mathbf{k}})^2+\Delta_0^2}}
\end{equation}

\noindent where $V_c$ is the strength of the coupling, and $n_0$ is the total number of quasiparticles in the quantum dot. $n_0$ is around $500$ \cite{Cheng:2015aa}. Notice that the factor $\xi_{\mathbf{q}}-E_{\mathbf{q}-\mathbf{k}}$ in fact is equivalent to the frequency of the electron in Green's function $\omega$. The calculation of the self-energy is presented in Appendix \ref{sec:cs}.

\section{even-odd effect within the boson-fermion model}
\label{sec:dos}
With Eqs. \ref{eq:gamma},\ref{eq:sigma}, the one-particle Green's function can be written as:
\begin{equation}
G(\mathbf{k},\omega)=\frac{1}{\omega-\epsilon_\mathbf{k}-\Sigma'- i \Gamma }
\end{equation}
where $\epsilon_\mathbf{k}=\sqrt{\Delta^2+(\frac{\hbar^2 (k^2-k_F^2)}{2 m^*})^2} \approx\sqrt{\Delta^2+(\hbar v^*_F (k-k_F))^2} $, $v^*_F\equiv \frac{\hbar k_F}{m^*}$, $\Sigma'=\Sigma'_\mathbf{k}(\omega)|_{k=k_F}$, $\Gamma=\Gamma_\mathbf{k}(\omega)|_{k=k_F}$. Numerical calculations show that $\Sigma' \ll \Delta$ at low temperature and consequently $\Sigma'$is negligible.

The density of states is

\begin{equation}
\label{eq:dos}
D(\omega)=\int_{-\infty}^\infty L A(\mathbf{k},w) d \mathbf{k}
\end{equation}
where $A(\mathbf{k},\omega)=-\frac{1}{\pi}{\rm Im}(G(\mathbf{k},\omega))$ and $L\approx 500{\rm nm}$ is the length of the small island in the middle of the single electron transistor. Notice that only $\epsilon_{\mathbf{k}}$ is dependent on the momentum $\mathbf{k}$ in the spectral weight function $A(\mathbf{k}, \omega)$. This allows us to deduce an exact result in the mathematical expression of $D(\omega)$ with the residue theorem (see Appendix \ref{sec:ap}).

As shown in Fig. \ref{bfdos}, the decay and formation of the electron pairs produces many gap states and broadens the van Hove singularity in the density of states. The effects can reduce the even/odd free energy difference $\delta {\cal F}_{\rm e/ o}$, and the effective excitation number for the unpaired electron in the odd parity state $N_{\rm eff}$, and this can be measured in experiments. In addition, zero superconducting gap makes a finite spectral function at Fermi level, and in that case, the density of states, $D(\omega)$, is finite at the Fermi level. This can destroy the even-odd effects. As shown in Fig. \ref{bfdos}, the results on the even-odd effects calculated by the boson-fermion model is very similar to those we obtained from Dynes' model density of states.

\begin{figure}
  \includegraphics[width=240pt]{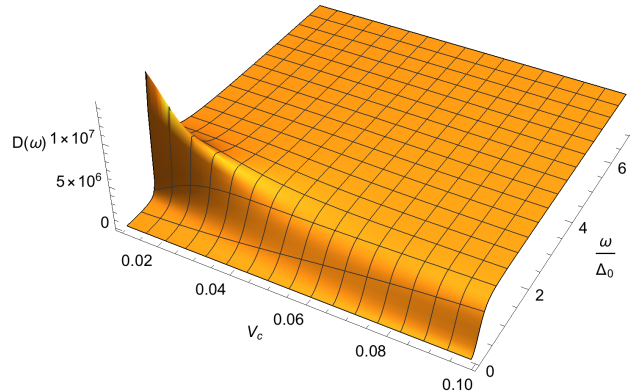}
  \caption{The density of states, $D(\omega)$, is plotted along with the variation of $V_c$. $D(\omega)$ and $V_c$ are plotted in atomic units.}
  \label{bfdos}
\end{figure}

\begin{figure}
  \includegraphics[width=250pt]{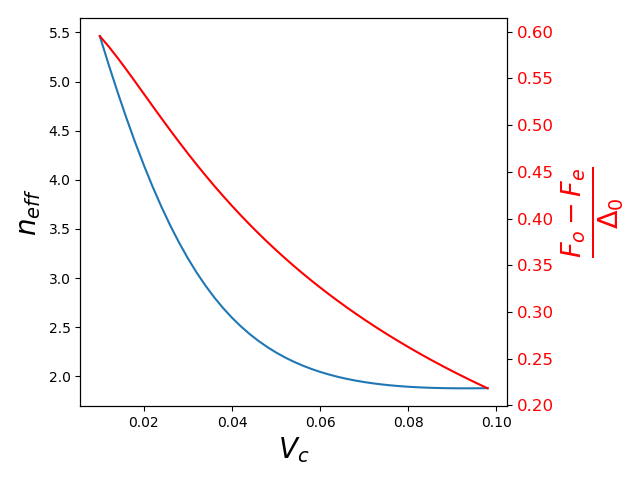}
  \caption{The even-odd free energy difference and the effective excitation number for the unpaired electron versus $V_c$. It is plotted in atomic units. $m^*= m_e$, $n_0=500$, $L=530 {\rm nm}$, $v^*_F=8.8 \times 10^3 {\rm m/s}$ and $v=0.073c$,where $c$ is the speed of light in vacuum.}
  \label{bf}
\end{figure}

\section{Conclusion}
\label{sec:con}
In the present paper, we argue that the even-odd effects seen in the normal state of STO originate from superconducting preformed pairing. This in turn imposes severe constraints on the density of states and consequently any microscopic model aimed at explaining the superconducting and normal state of STO/LAO. First, the density of states at Fermi level should be zero below and above $T_c$. A superconducting gap is required for the electron system to demonstrate even-odd effects above $T_c$. Next, the gap states are necessary to reduce the even/odd free energy difference and weaken the even-odd effects. Finally, the van Hove singularity needs to be broadened out in order to obtain a small $N_{\rm eff}$. These constraints for the density of states are not immediately satisfied by most current microscopic theories. On the other hand, the theories are successful in explaining the microscopic origin of the superconducting gap $\Delta$ and the dispersion of Bogoliubov quasiparticles.

The broadening of the van Hove singularity in the density of states may be a fingerprint of the lifetime effects of electron pairs. The signals in the single electron transistor experiments are very sensitive in detecting the lifetime effects, as well as the existence of the superconducting gap, the gap states and the broadening of the van Hove singularity. Moreover, compared to the experimental condition of scanning tunneling spectroscopy, that of the single electron devices can be relatively more easily satisfied in some strongly interacting electron systems. Furthermore, the decay and formation of electron pairs may widely exist in many different types of superconductors, including BCS superconductors. The application of the single electron transistor devices to study novel superconductors is therefore very promising.

The microscopic origin of phenomenological interaction potential is still unknown. Phonon-electron interactions, electron-electron interactions, etc. may participate in the actual microscopic mechanism\cite{xing}. Further theoretical and experimental investigations are needed to elucidate the detailed microscopic model of the superconducting and normal state of SrTiO\textsubscript{3}. 

\begin{acknowledgments}
One of us (X. Y.) gratefully acknowledges the support from China Scholarship Council under Grant No. 201506130054. We thank Peter B. Littlewood, Alexander Edelman, Anthony Ruth and Xiaoyu Ma for helpful discussions. This work is financially supported by National Natural Science Foundation of China under Grant No. 11675051.
\end{acknowledgments}
\appendix

\section{Calculations of the self-energy}
\label{sec:cs}
The decay rate of quasiparticles is defined to be $ \Gamma = \mathrm{Im}(\Sigma(\mathbf{k}, \omega))$. In order to facilitate the calculations, it is assumed that $\Sigma(\mathbf{k},\omega)=\Sigma(\mathbf{k},\omega) \big|_{|\mathbf{k}|=k_F}$. From Eq. \ref{eq:sigma} and the above approximations, the decay rate of electron pairs is

\begin{equation}
\begin{split}
&\Gamma (\omega) = \frac{1}{\hbar^2}\int \frac{L |V_1(\mathbf{q})|^2d \mathbf{q}}{2 n_0} \delta(\omega-\frac{(\xi_{\mathbf{q}}-E_{\mathbf{q})}}{\hbar})\\
&\times (\frac{1}{e^{\beta \xi_{\mathbf{q}}}-1}+\frac{1}{e^{\beta E_{\mathbf{q}}}+1}).\\
\end{split}
\end{equation}

When $q$ is very small and $\hbar v^*_F q \ll \Delta$, the equation $\omega-\frac{(\xi_{\mathbf{q}}-E_{\mathbf{q})}}{\hbar}=0$ becomes $\omega-(\xi_0+v|q|-\mu_b)+\Delta\approx 0$ where $\hbar$ is assumed to be unity. The chemical potential of bosons is set to be $\xi_0-\mu_b \approx \Delta$. This leads to the solution that $\omega \approx v|q|$. The frequency of the electrons $\omega$ is in the order of $\sim \Delta$. This self-consistently proves that $q$ is very small and $\hbar v^*_F q \ll \Delta$. With the results, we obtain

\begin{equation}
\label{eq:gamma}
\Gamma (\omega) \approx \frac{L |V_1(\frac{\omega}{v})|^2}{v n_0} (\frac{1}{e^{\beta (\Delta+\omega)}-1}+\frac{1}{e^{\beta \Delta}+1}).
\end{equation}

With the particle-hole parity symmetry and Kramers-Kr\"{o}nig relation, the real part of the self-energy is
\begin{equation}
\label{eq:sigma}
\Sigma'(\omega)=\mathrm{Re}(\Sigma(k=k_F, \omega)) = \frac{2 \omega}{\pi}P\int_0^\infty \frac{\Gamma(\omega')}{\omega'^2-\omega^2} d \omega'
\end{equation}

Numerical calculations show that the real part of the self-energy has a negligible effect in generating the predicted density of states.

\section{Calculations of the Density of States}
\label{sec:ap}
The denominator of the spectral weight function is $(w-\sqrt{\Delta^2+(\hbar v^*_F (k-k_F))^2}-\Sigma')^2+\Gamma^2$. If the denominator equals to zero, there are four solutions of the momentum $k$ that two solutions are in upper half-plane of the complex plane of $k$ and two solutions are in the lower half. Furthermore, there are four different cases, if we set $a=(\omega-\Sigma')^2-\Gamma^2-\Delta^2, b=2 \Gamma (\omega-\Sigma')$

  \[
                \left\{
                \begin{aligned}
                  a>0, b>0...................\romannum{1}~~\\
                  a>0, b<0...................\romannum{2}~\\
                  a<0, b>0...................\romannum{3}\\
                  a<0, b<0...................\romannum{4}.
                \end{aligned}
              \right.
  \]

For case \romannum{1}, four solutions of the momentum $k$ in the upper half-plane are

                \begin{align}
                  k_1 &= k_F - \frac{1}{v^*_F}\sqrt{R}e^{\frac{-\theta i}{2}}\\
                   k_2 &=k_F + \frac{1}{v^*_F}\sqrt{R}e^{\frac{\theta i}{2}}
                \end{align}

where $\theta=Arctan\frac{b}{a}, R=\sqrt{a^2+b^2}$. For case \romannum{2},

                \begin{align}
                   k_1=k_F + \frac{1}{v^*_F}\sqrt{R}e^{\frac{-\theta i}{2}}\\
                   k_2=k_F - \frac{1}{v^*_F}\sqrt{R}e^{\frac{\theta i}{2}}.
                \end{align}

For case \romannum{3},

                \begin{align}
                   k_1=k_F - \frac{1}{v^*_F}\sqrt{R}e^{\frac{-(\theta+\pi) i}{2}}\\
                   k_2=k_F + \frac{1}{v^*_F}\sqrt{R}e^{\frac{(\theta+\pi) i}{2}}.
                 \end{align}

For case \romannum{4},

                \begin{align}
                   k_1=k_F - \frac{1}{v^*_F}\sqrt{R}e^{\frac{(\theta-\pi) i}{2}}\\
                   k_2=k_F + \frac{1}{v^*_F}\sqrt{R}e^{\frac{(-\theta+\pi) i}{2}}.
                  \end{align}

  After applying Jordan's lemma and the residue theorem, we obtain the density of states, for case \romannum{1},

\begin{equation}
D(\omega)=\frac{2L}{v^*_F}\bigg[\frac{\omega - \Sigma'}{\sqrt{R}}\cos \bigg (\frac{\theta}{2}\bigg)+\frac{\Gamma}{\sqrt{R}}\sin\bigg(\frac{\theta}{2}\bigg)\bigg].
\end{equation}

For case \romannum{2},
\begin{equation}
D(\omega)=-\frac{2L}{v^*_F}\bigg[\frac{\omega - \Sigma'}{\sqrt{R}}\cos\bigg(\frac{\theta}{2}\bigg)+\frac{\Gamma}{\sqrt{R}}\sin\bigg(\frac{\theta}{2}\bigg)\bigg].
\end{equation}

For case \romannum{3},
\begin{equation}
D(\omega)=\frac{2L}{v^*_F}\bigg[\frac{\omega - \Sigma'}{\sqrt{R}}\cos \bigg(\frac{\theta+\pi}{2}\bigg)+\frac{\Gamma}{\sqrt{R}}\sin\bigg(\frac{\theta+\pi}{2}\bigg)\bigg].
\end{equation}

For case \romannum{4},
\begin{equation}
D(\omega)=-\frac{2L}{v^*_F}\bigg[\frac{\omega - \Sigma'}{\sqrt{R}}\cos\bigg(\frac{\theta-\pi}{2}\bigg)+\frac{\Gamma}{\sqrt{R}}\sin\bigg(\frac{\theta-\pi}{2}\bigg)\bigg].
\end{equation}

\bibliography{draft4}
\bibliographystyle{apsrev4-1}

\end{document}